\input{aipcheck}

\documentclass[
    ,final            
  ]
  {aipproc}

\layoutstyle{6x9}

\begin{document}

\title{The value of density measurements in stellar coronae}


\author{Jan-Uwe Ness}{
  address={Department of Physics, Rudolf Peierls Centre for
Theoretical Physics, University of Oxford, 1 Keble
Road, Oxford OX1\,3NP}
}

\author{Carole Jordan}{
  address={Department of Physics, Rudolf Peierls Centre for
Theoretical Physics, University of Oxford, 1 Keble
Road, Oxford OX1\,3NP}
}

\begin{abstract}
The grating instruments on board Chandra and XMM-Newton now allow
measurements of electron densities. These rely on the ratios of fluxes
in emission lines, where one line depends on both collisional and
radiative decay rates. The electron density is required to constrain
the physical extent of the emitting region, and large samples of
measurements are of interest in the context of trends in coronal
activity. Here we discuss the important He\,{\sc i}-like ions and the
differences in densities that result when different current data bases
are used.
\end{abstract}

\maketitle


\section{Requirements for measuring plasma densities}

\begin{figure}[!ht]
 \rotatebox{90}{\includegraphics[height=.3\textheight]{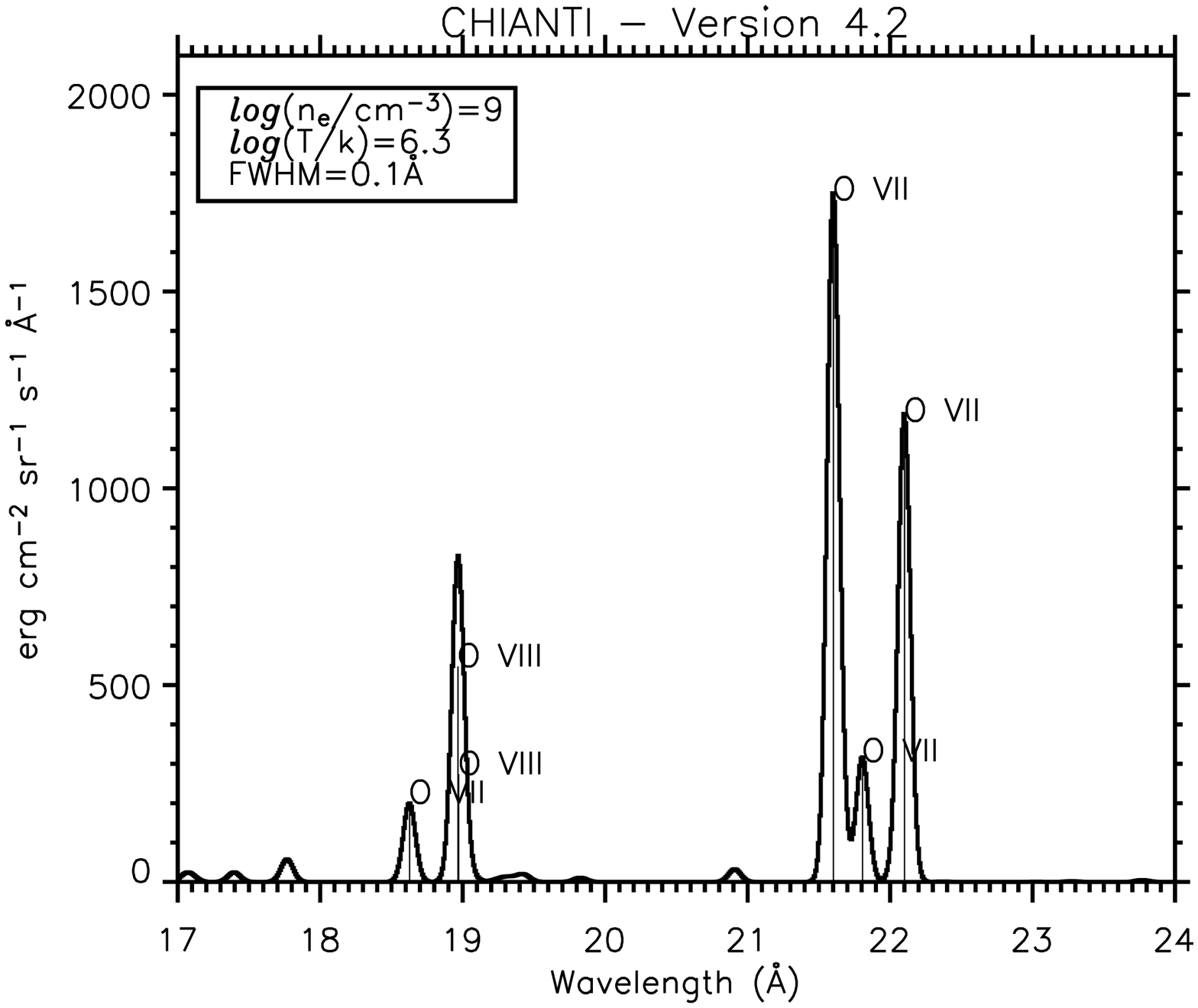}}
 \rotatebox{90}{\includegraphics[height=.3\textheight]{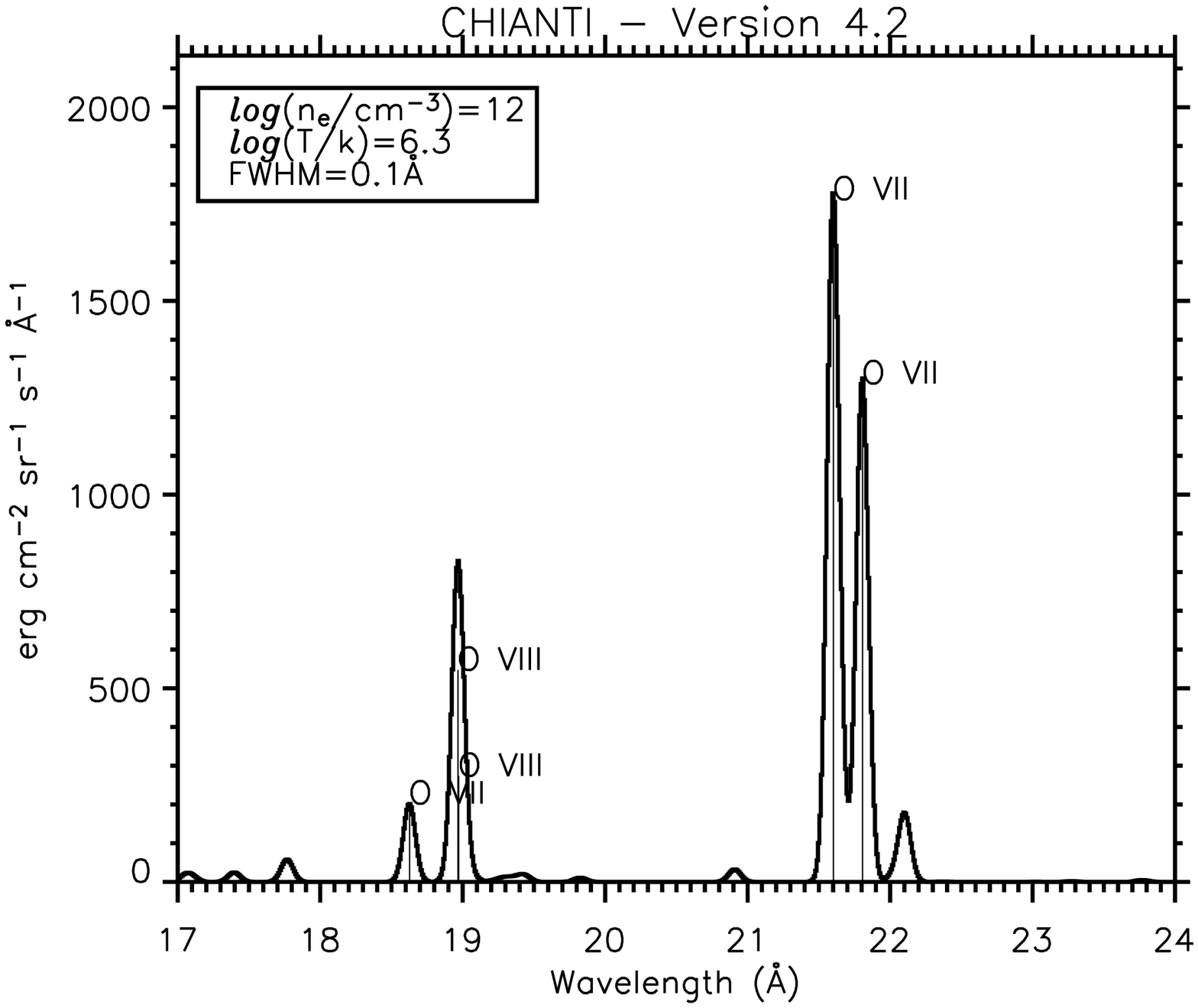}}
\caption{\label{hres}Simulations of spectra with low densities (left) and high densities.}
\end{figure}

 The spectrum emitted by an X-ray plasma reflects the physical conditions of the plasma,
especially the plasma temperature and the plasma density. While different plasma
temperatures affect an X-ray spectrum quite strongly (broad band continuum emission
and appearance and disappearance of strong lines originating from ions in different
ionization stages), the effects of different densities are rather subtle.
This is demonstrated with Fig.~\ref{hres}, where two simulated spectra are shown. For
simplicity we chose an isothermal plasma in one case with a high density and in the other
with a low density. It can be seen that differences are only detectable in a few
emission lines, and therefore density measurements in X-ray plasmas can only be
carried out with high-resolution spectroscopy.

\section{Density measurements with He-like triplets}

The He-like ions provide a valuable method of measuring the electron
density ($n_e$), in principle over a wide range of electron temperatures
($T_e$). The method \citep[see][]{gj69} depends on the competition between collisions
between the 1s2s\,$^3$S and 1s2p\,$^3$P levels and the radiative decay in
the forbidden line (1s2s\,$^3$S -- 1s$^2\,^1$S). The ratio of the flux
in the forbidden line to that in the intersystem (+magnetic
quadrupole) line, then becomes sensitive to $n_e$, and can be
parameterized as $f/i=R_0/(1+n_e/N_c)$,
where $R_0$ is the ratio in the low density limit and $N_c$ is the
critical density. Both $R_0$ and $N_c$ contain atomic data relating to
collisional and radiative rates. Fig.~\ref{hres} shows
simulated spectra for O\,{\sc vii} at densities of $10^9$\,cm$^{-3}$ and
$10^{12}$\,cm$^{-3}$, between which the ratio depends on $n_e$.

\begin{figure}
  \includegraphics[height=.25\textheight]{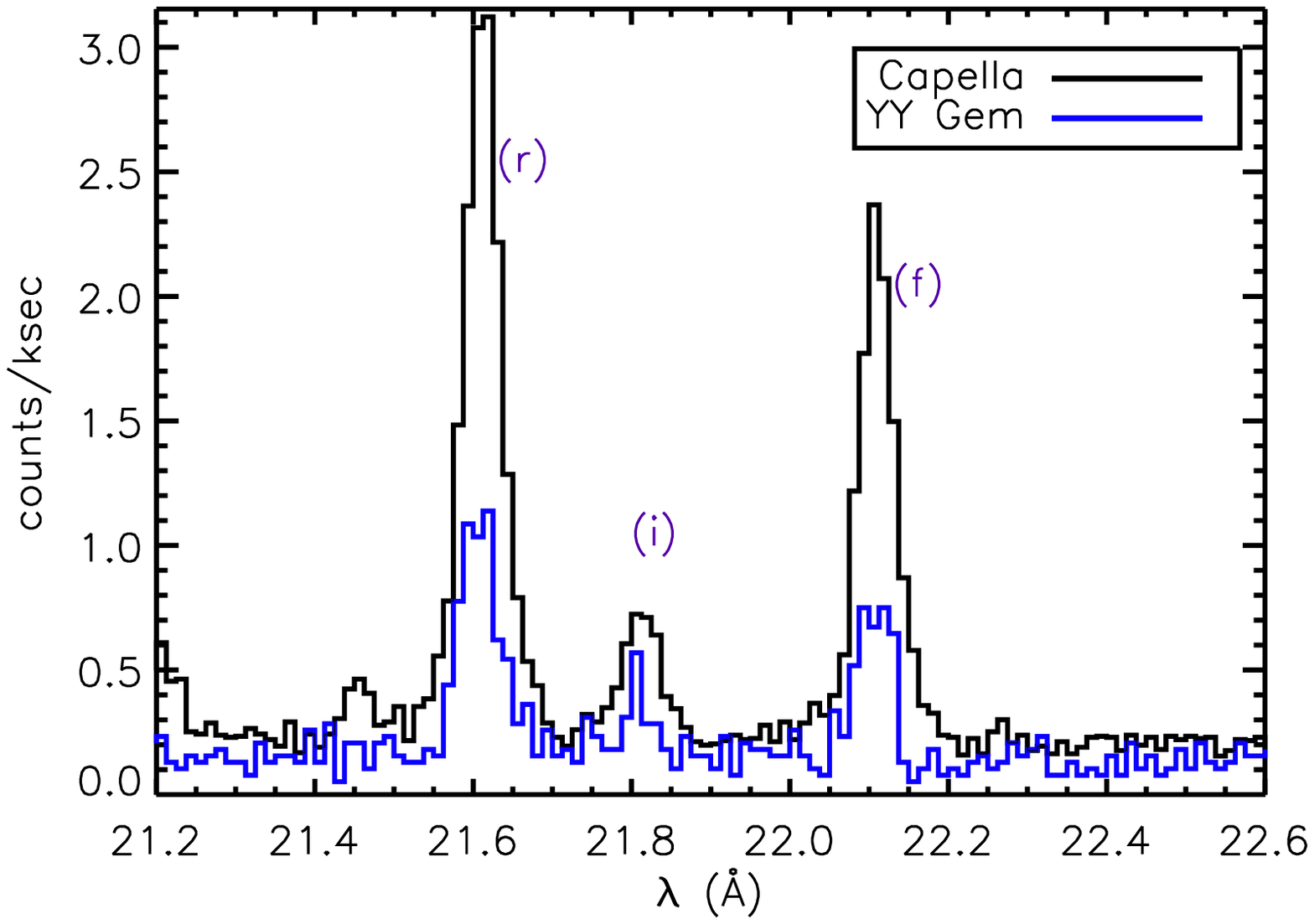}
  \includegraphics[height=.25\textheight]{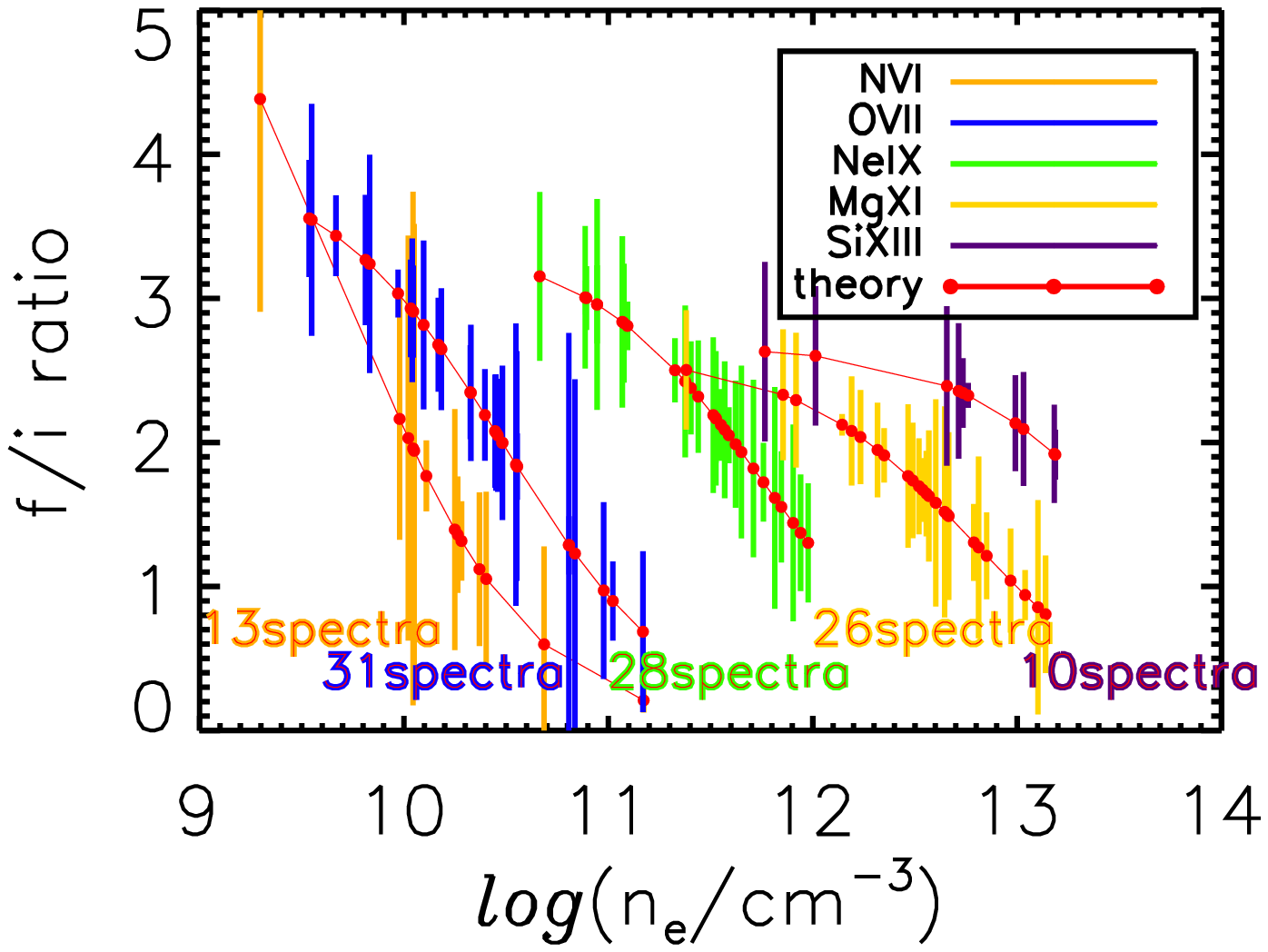}
  \caption{\label{meas}{\bf Left:} Chandra LETGS measurements of the
O\,{\sc vii} He-like triplets for Capella and for YY\,Gem (blue), scaled to the
distance of Capella. The spectra show different behaviour in the ratio of the lines
marked with (i) and (f). {\bf Right:} Density measurement with different
He\,{\sc i}-like ions for a large sample of stellar coronae. Error bars
(1$\sigma$ statistical errors)
are large due to faint intercombination lines in the case of large f/i
ratios. Mg\,{\sc xi} and Si\,{\sc xiii} cannot constrain densities below $10^{12.5}$\,cm$^{-3}$.}
\end{figure}

The Chandra LETGS measurements of Capella and YY\,Gem (Fig.~\ref{meas}) show
clear differences in f/i ratios, which indicate different densities in the two
coronae.

\begin{figure}
 \includegraphics[height=.25\textheight]{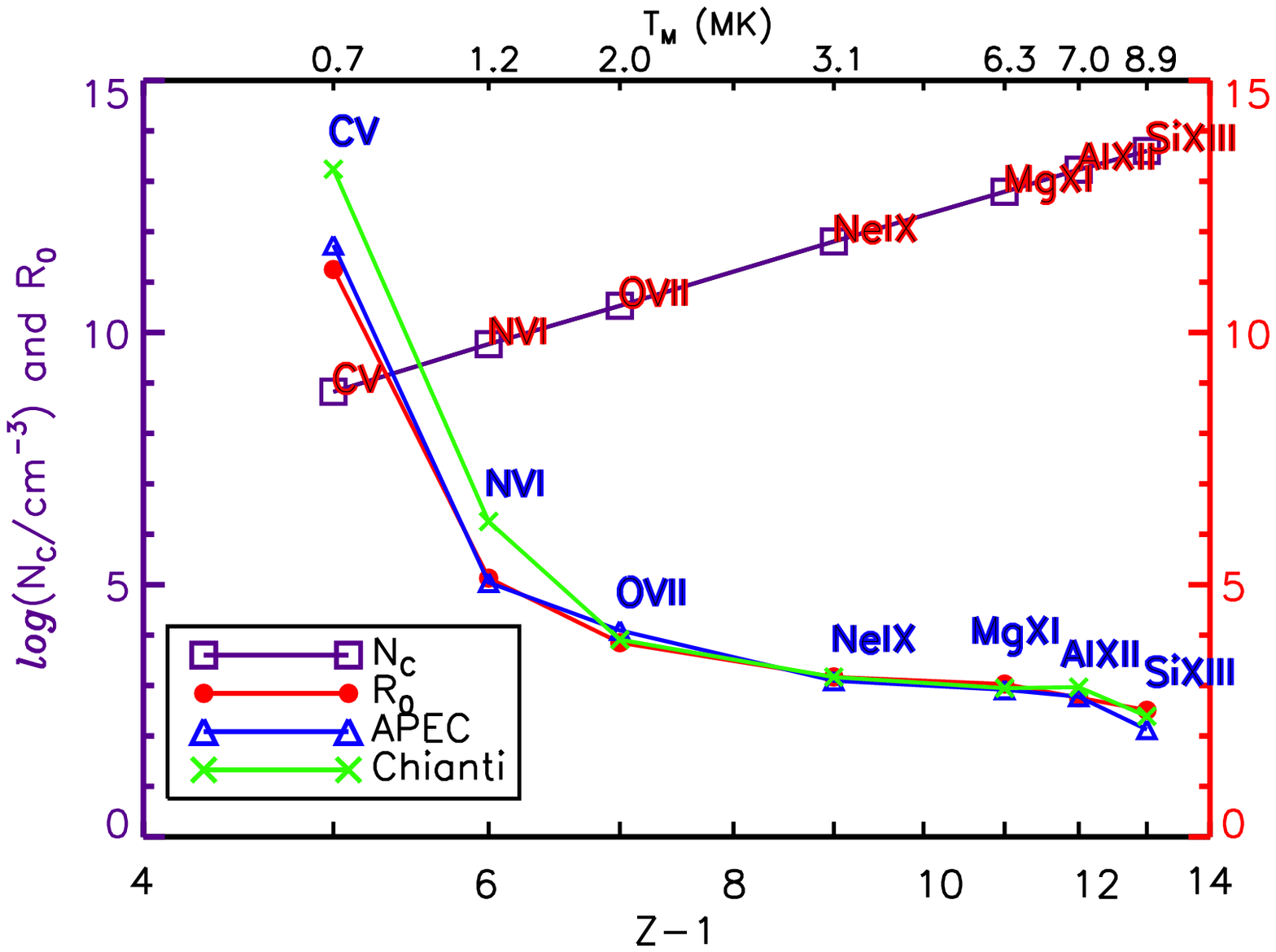}
 \includegraphics[height=.25\textheight]{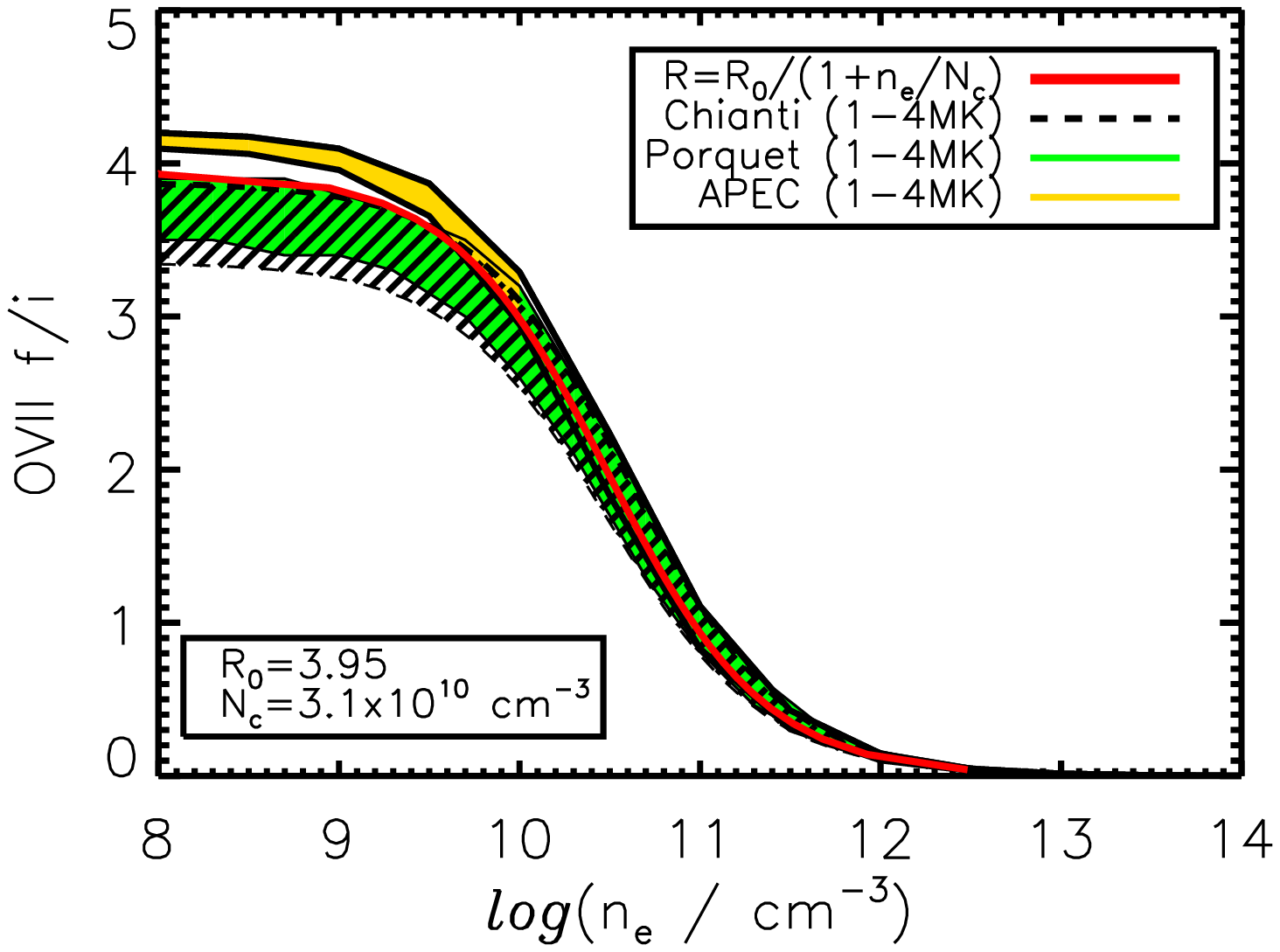}
\caption{\label{he}Left panel: comparison of $R_0$ (from different data
bases) and $N_c$ \citep[][ in purple]{blum72}.
 Right panel: density-dependent f/i line flux ratios for O\,{\sc vii}.}
\end{figure}

The atomic rates have different dependences on Z - 1, where Z is the
nuclear charge, but should depend smoothly on Z - 1, for the ions of
interest. Fig.~\ref{he} (left) shows the parameters $R_0$ and $N_c$ as a
function of Z - 1. This brings out the differences between the data bases
CHIANTI and APEC and shows $\log N_c$ from \cite{blum72}.
Values of $R_0$ are indicated by different colours;
this plot suggests that the data in $R_0$ for C\,{\sc v} and N\,{\sc vi}
in CHIANTI need to be re-examined, and that those for Si\,{\sc xiii} warrant closer
examination. Note that $\log N_c$ depends linearly on Z-1.

The right hand panel of Fig.~\ref{he} shows the resulting variation of the ratio f/i
for O\,{\sc vii}, indicating the importance of accurate atomic data when observed
ratios are close to the low density limit. Measured densities will always be
averages, but the He-like triplets formed at high temperatures
probe only high densities (above $10^{12}$\,cm$^{-3}$), while low-temperature ions
measure only lower densities ($\sim 10^{10}-10^{11}$\,cm$^{-3}$); this leaves two cases
unexplored: low densities in hot plasma and high densities in cool plasma.

\section{What can we learn from density measurements?}

 In stellar coronae the density measurements provide an important link between
physical and geometrical properties, through the Emission Measure
EM\,$= n_e^2 \times V$ ($n_e$: electron density and $V$: emitting volume).
Densities are also needed to explore the optical depths of the lines:
$\tau\propto \int n_ed\ell$, where $\ell$ is the path length (cm), and hence
emitting areas.

\section{Results and Conclusions}

 A survey of density measurements was carried out by \cite{denspaper}.
Densities were measured from the He-like triplets (O\,{\sc vii} and Ne\,{\sc ix})
probing densities at temperatures $2\times 10^6$\,K and at $4\times 10^6$\,K.
Measurements from higher-Z ions (Mg\,{\sc xi} and Si\,{\sc xiii}) result only in
low-density limits $<10^{12.5}$\,cm$^{-3}$ (Fig.\ref{meas}, right).
The observed range of f/i ratios shows that real differences in $n_e$ do occur
between different coronae.

Accurate atomic data are essential in order to use these and other
diagnostics of $n_e$, thus exploiting the excellent new spectra being obtained.
Values of $n_e$ can then be used to constrain physical models of the
atmosphere.


\begin{theacknowledgments}
JUN acknowledges support from PPARC under grant number PPA/G/S/2003/00091
\end{theacknowledgments}



\bibliographystyle{aipproc}   

\bibliography{jn,astron}

\IfFileExists{\jobname.bbl}{}
 {\typeout{}
  \typeout{******************************************}
  \typeout{** Please run "bibtex \jobname" to optain}
  \typeout{** the bibliography and then re-run LaTeX}
  \typeout{** twice to fix the references!}
  \typeout{******************************************}
  \typeout{}
 }

\end{document}